\documentclass[a4paper,onecolumn]{IEEEtran}

\usepackage{amsmath,amssymb,amsfonts}
\usepackage{mathrsfs}
\usepackage{graphicx}
\usepackage{tabularx}
\usepackage{subcaption}
\usepackage{textcomp}
\usepackage{smartdiagram}
\usepackage{adjustbox}
\usepackage{pgfplots}
\usepackage{float}
\usepackage{xcolor}
\usepackage{colortbl}
\usepackage{bm}
\usepackage[boxed,commentsnumbered, linesnumbered,ruled,vlined]{algorithm2e}
\usepackage[sorting=none]{biblatex}
\def\BibTeX{{\rm B\kern-.05em{\sc i\kern-.025em b}\kern-.08em
    T\kern-.1667em\lower.7ex\hbox{E}\kern-.125emX}}
\usepackage{epstopdf}
\usepackage{siunitx}
\usepackage{tikz}
\usepackage{circuitikz}
\usepackage{dirtytalk}
\usepackage{svg}
\usepackage{listings}
\usetikzlibrary{shapes.geometric, arrows, positioning}
\usepackage{titlesec}
\titlespacing{\subsubsection}{0pt}{10pt}{5pt}

\usepackage[left=2.2cm, right=2.2cm, top=3.5cm, bottom=2.5cm]{geometry}

\usepackage[skip=5pt plus1pt, indent=10pt]{parskip}

\usepackage{fancyhdr}
\graphicspath{{./figures/}}

\tikzstyle{startstop} = [rectangle, rounded corners, minimum width=3cm, minimum height=1cm,text centered, draw=black, fill=red!30]
\tikzstyle{io} = [trapezium, trapezium left angle=70, trapezium right angle=110, minimum width=3cm, minimum height=1cm, text centered, draw=black, fill=blue!30]
\tikzstyle{process} = [rectangle, minimum width=3cm, minimum height=1cm, text centered, text width=4cm, draw=black, fill=orange!30]
\tikzstyle{decision} = [diamond, minimum width=3cm, minimum height=1cm, text centered, draw=black, fill=green!30]
\tikzstyle{arrow} = [thick,->,>=stealth]

\begin{document}

\title{Pragmatic Formal Verification Methodology for Clock Domain Crossing (CDC)}

\author{
\IEEEauthorblockA{\vspace{4mm}Aman Kumar,
Infineon Technologies,
Dresden, Germany
(\textit{aman.kumar@infineon.com})}

\IEEEauthorblockA{Muhammad Ul Haque Khan,
Cadence Design Systems,
Munich, Germany
(\textit{muhammad@cadence.com)})}

\IEEEauthorblockA{Bijitendra Mittra,
Cadence Design Systems,
Bangalore, India
(\textit{bijitm@cadence.com})}}

\maketitle

\thispagestyle{fancy}


\begin{abstract}
\textbf{\emph{Abstract}\!
\textemdash Modern System-on-Chip (SoC) designs are becoming more and more complex due to the technology upscaling. SoC designs often operate on multiple asynchronous clock domains, further adding to the complexity of the overall design. To make the devices power efficient, designers take a Globally-Asynchronous Locally-Synchronous (GALS) approach that creates multiple asynchronous domains. These Clock Domain Crossings (CDC) are prone to metastability effects, and functional verification of such CDC is very important to ensure that no bug escapes. Conventional verification methods, such as register transfer level (RTL) simulations and static timing analysis, are not enough to address these CDC issues, which may lead to verification gaps. Additionally, identifying these CDC-related bugs is very time-consuming and is one of the most common reasons for costly silicon re-spins \cite{VerStudy}. This paper is focused on the development of a pragmatic formal verification methodology to minimize the CDC issues by exercising Metastability Injection (MSI) in different CDC paths.}
\end{abstract}


\section{Introduction}
\label{sec:intro}
\begin{figure}[h!]
\begin{tikzpicture}
    \begin{axis}[
            ybar,
            bar width=.12cm,
            width=\textwidth,
            height=.5\textwidth,
            legend style={at={(0.5,1)},
                anchor=north,legend columns=-1},
            symbolic x coords={Logical/functional,Clocking,Analog,Crosstalk,Power consumption,Mixed-signal interface,Yield,Reliability,Timing -- Path too slow,Firmware,Timing -- Path too fast,IR drops,Safety,Security,Other},
            xtick={data},
            ymin=0,ymax=55,
            ylabel={\%},
            x tick label style={font=\footnotesize,rotate=45, anchor=east},
            xticklabels={Logical/functional,\textcolor{red}{Clocking},Analog,Crosstalk,Power consumption,Mixed-signal interface,Yield,Reliability,Timing -- Path too slow,Firmware,Timing -- Path too fast,IR drops,Safety,Security,Other}
        ]
        \addplot table[x=Issue,y=2016]{\mydata};
        \addplot table[x=Issue,y=2018]{\mydata};
        \addplot table[x=Issue,y=2020]{\mydata};
        \addplot table[x=Issue,y=2022]{\mydata};
        \legend{2016, 2018, 2020, 2022}
    \end{axis}
\end{tikzpicture}
\caption{Type of ASIC flaws contributing to re-spin \cite{VerStudy}}
\label{respin}
\end{figure}
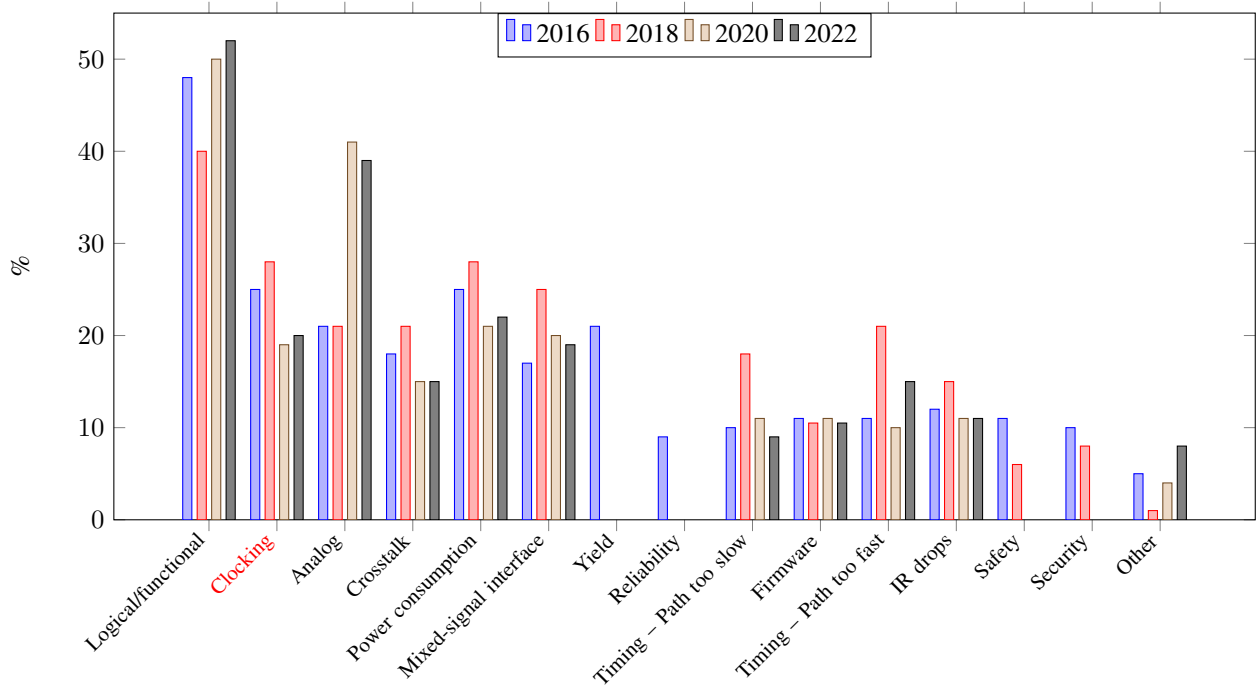

Integrated Circuits (ICs) are a fundamental component that is essential for modern electronic devices. They can perform a wide range of functions, from signal processing to artificial intelligence, and are used in many industries. However, as ICs have become more complex, verification of their designs has become a major challenge \cite{VerStudy}. In fact, design verification now takes up a significant amount of project time and failures to identify all functional issues can result in delays and even setbacks in terms of expensive silicon re-spin. A recent study conducted by Siemens and Wilson Research Group shows that design verification consumes approximately \SI{60}{\percent} of the total project time \cite{VerStudy}. Given the fierce competition in the semiconductor market and ongoing problems like chip shortages, a single silicon re-spin could cost a lot of money and prolong the product’s time to market by months, significantly decreasing the chip’s market share and profit potential. In the study \cite{VerStudy} from Fig.~\ref{respin}, flaws in clocking are the third largest contributor towards the re-spin. One particular challenge is CDC, which occurs when signals cross asynchronous clock domains in intricate IC designs. To ensure that ICs are tolerant of these problems and function properly, new verification methods need to be devised.

\pgfplotstableread[row sep=\\,col sep=&]{
    Issue &	2016 &2018	&2020	&2022 \\
Logical/functional&	48&	40&	50&	52 \\
Clocking&	25	&28	&19	&20 \\
Analog&	21&	21&	41&	39 \\
Crosstalk&	18&	21&	15&	15 \\
Power consumption&	25&	28&	21&	22 \\
Mixed-signal interface&	17&	25&	20&	19 \\
Yield	&			21& & & \\
Reliability	&			9& & & \\
Timing -- Path too slow	&10&	18&	11&	9 \\
Firmware	&11&	10.50&	11&	10.50 \\
Timing -- Path too fast	&11&	21&	10&	15 \\
IR drops	&12&	15&	11	&11 \\
Safety		&	11&	6& & \\
Security	&		10&	8& & \\
Other&	5&	1&	4&	8 \\
    }\mydata

Furthermore, the GALS approach increases the overall CDC complexity of the design. The state-of-the-art verification techniques such as conventional RTL simulation or Static Timing Analysis (STA) alone are not enough to address the CDC issues of such intricate SoC designs. Hence, it is essential to devise a new verification methodology to address the CDC issues of such complex designs. The new CDC verification methodology should ensure that the Design Under Verification (DUV) is tolerant to problems such as metastability effects, re-convergence errors, Reset Domain Crossing (RDC) issues, and data loss faults. This basically implies that in the design, the CDC signals can span across asynchronous clock domains without being missed or wrongly sampled, thereby preventing metastable values from propagating further downstream.

\section{Technical Background}
A Clock Domain (CD) is a part of the design that is driven by one or more clocks that are synchronous to each other (i.e., active edges are aligned). However, most real-life designs operate on multiple asynchronous clocks. A CD boundary arises when the clocks change. There are several reasons for using multiple CDs \cite{arm_book}:

\begin{itemize}
    \item Power and performance folding: High-frequency logic often requires more power than lower frequency logic. Therefore, designs often have a larger amount of lower-speed logic than faster-clocked logic. The design is more energy efficient while operating parts of a circuit with a lower frequency clock than the other part of the design. In a low-power design, one can turn off one CD while keeping other CDs active, thus reducing power consumption.
    \item Physically separate clocks: Usually each system has its own clock generator to make sure that they work unanimously even if networking cables, such as Ethernet, are disconnected. Quartz-based crystal oscillators generally used as clocks will differ in accuracy with temperature, supply voltage, and crystal age. However, this creates different clock domains at the receiver end.
\end{itemize}

There are several issues associated with designs that have multiple CDCs. These problems are well-documented in \cite{cdc_issues} including \cite{mark_cdc}:

\begin{itemize}
    \item Setup and hold time violations due to metastability
    \item Functional errors due to convergence of synchronized signals
    \item Functional errors due to divergence through multiple synchronizers
\end{itemize}

\subsection{Metastability}
Metastability is a state of instability in a circuit where a flip-flop output cannot settle to a stable '0' or '1' logic level within the required time to function correctly. This can result in intermediate voltage levels being processed incorrectly and the circuit remaining in an unstable condition for a period of time, resulting in functional failures.

\subsection{Causes of Metastability}
\begin{figure}[h!]
\centering
  \includegraphics [width=0.7\textwidth] {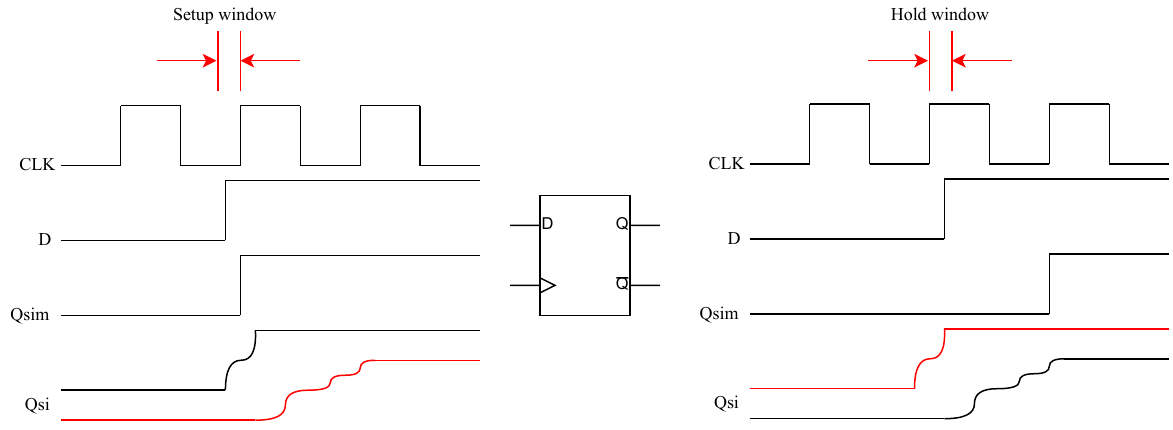}
\caption{Setup and hold time violation \cite{cadence_cdc}}
\label{violation}
\end{figure}

Metastable behaviour occurs in flip-flops due to setup and hold timing violations. The setup time is the minimum amount of time during which the input data must arrive and be stable before the active edge of the clock. When the input data changes during the setup time window, a setup violation occurs. The hold time is the minimum amount of time during which the input data must be stable after the active edge of the clock. When the input data changes during the hold time window, a hold violation is observed. When a setup or hold timing violation occurs, the output of the flip-flop may go metastable as shown in Fig.~\ref{violation}. For CDC signals coming from asynchronous clock domains, it is not possible to prevent metastability from occurring due to the non-deterministic clock relationships.

Setup/hold violations cannot usually be detected in an RTL simulation. When there is a setup violation as depicted in Fig.~\ref{violation}, the simulation always captures a '1', while in silicon it may produce a '0' or '1'. As a result, during setup violation, the output transition of the flip-flop may get delayed by a cycle in silicon. When there is a hold violation as depicted in the above figure, the simulation always captures a ‘0’, whereas in silicon it may produce either ‘0’ or ‘1’. During hold violation, the output transition of the flip-flop may occur one cycle early in the silicon. A flip-flop becomes unstable when setup or hold conditions are violated, and it eventually settles down to either ‘0’ or ‘1’ logic level, after an unpredictable delay.

\subsection{Effects of Metastability}
The metastability effect can have the following consequences:

\begin{itemize}
    \item It can lead to unanticipated system behaviour or functional failures.
    \item Propagation of metastable data (unstable data) to different blocks of the SoC design can induce high currents and can eventually result in chip burnout.
    \item The design may enter into an unknown state or into a deadlock scenario due to the metastable behaviour of the control signals.
\end{itemize}

In multi-clock chip architectures, metastability cannot be avoided, but its detrimental effects can be mitigated by using the right methods, such as the use of synchronizers for clock domain crossing signals \cite{sunburst_cdc}. To minimize issues related to metastability propagation and data loss, different analysis such as structural CDC analysis, verification of the CDC protocol and functional CDC analysis are performed \cite{mark_cdc}.

\section{Conventional CDC Verification Methodology}
\label{sec:conventional}
\begin{figure}[h!]
\centering
  \includegraphics [width=0.7\textwidth] {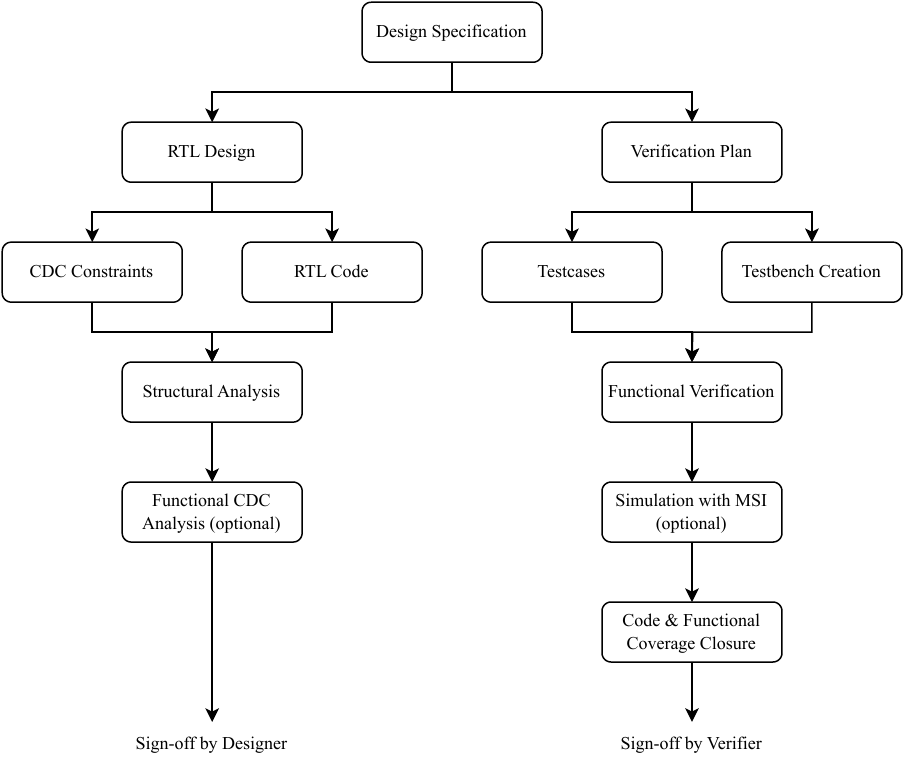}
\caption{Conventional CDC verification methodology}
\label{conv_method}
\end{figure}

In the conventional CDC verification process, as depicted in Fig.~\ref{conv_method}, design and verification are treated separately. Once the design specification is complete, the design engineer creates the RTL implementation of the code, which is then checked for structural clock domain crossover issues using a static CDC tool. The analysis looks for problems such as missing or incorrectly placed synchronizers and glitches caused by combinatorial logic on the synchronizer paths. To facilitate this analysis, the designer writes a series of CDC constraints that may include specifying CDC false paths, identifying static and constant signals, and so on. Once the structural analysis is completed, the designers can optionally perform a functional analysis to ensure that the synchronizer protocols are correct. This is the standard CDC sign-off process used by designers.

The work of a functional verification engineer starts with the creation of a test plan that aligns with the design specification. Subsequently, the engineer creates a simulation testbench and develops test cases to validate the functionality of the design. If all the test cases of the regression suite are passing, the verification engineer may choose to run the simulation using randomizable synchronizer models, although this scenario is not typical during the pre-silicon functional verification stage due to the limited time budget. Finally, the verification engineer declares the CDC verification sign-off when both code and functional coverage have reached \SI{100}{\percent}. However, there are a few drawbacks to the conventional verification approach. The gaps associated with the CDC sign-off flow of designers are:

\begin{itemize}
    \item When the design engineer writes CDC constraints manually, it is error-prone, and a wrong CDC constraint or assumption could lead to an incorrect structural analysis by the static CDC analysis tool. This may let a functional bug slip into the post-silicon phase, resulting in expensive silicon re-spins.
    \item There is no functional analysis for user-defined synchronizer schemes.
    \item Absence of functional analysis with metastability modelling required to detect issues such as metastability propagation and re-convergence errors.
\end{itemize}

There are also gaps associated with the CDC sign-off flow of verification engineers. The verification engineer usually does not perform formal functional verification with MSI, which is a significant reason why the functional bug escapes to the silicon. The second major gap is the use of inaccurate MSI models in simulation runs. Most standard MSI models only mimic setup violations, which are incomplete and insufficient to detect all CDC issues. Third, no coverage model has been defined to measure the completeness of the CDC verification, which is crucial to determine the sign-off quality of the DUV. The proposed verification methodology in this paper uses a combination of both formal (static) and simulation (dynamic) based verification methods to address the shortcomings of conventional CDC verification methodologies.

\section{Proposed CDC Verification Methodology}
\label{sec:proposed}
The drawbacks of conventional CDC verification methodology clearly highlight that there is a missing link between CDC sign-off and functional verification sign-off flows that usually run as independent activities, creating verification gaps. Here, the verification gaps are created. Our proposed flow addresses these gaps efficiently and effectively.

\begin{figure}[h!]
\centering
  \includegraphics [width=0.7\textwidth] {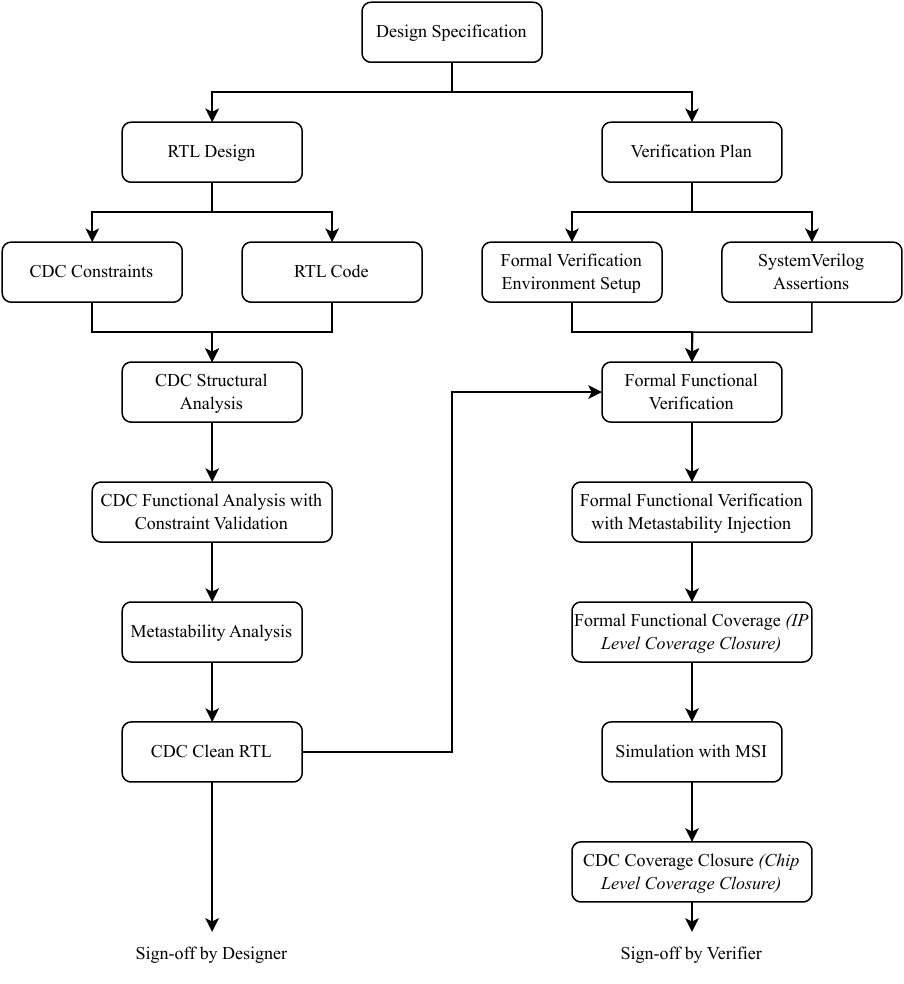}
\caption{Proposed CDC verification methodology \cite{binil}}
\label{methodology_flow}
\end{figure}

To overcome the drawbacks, we propose a more pragmatic verification methodology based on formal approach as shown in Fig.~\ref{methodology_flow}. Formal verification has advantages over simulation based verification as highlighted in \cite{fv_book}, \cite{aman_dvcon_configvermet} and \cite{aman_dvcon_ecc}. Although, there are associated limitations as well as mentioned in \cite{fv_book} and \cite{aman_dvcon_ecc} but with the right approach, formal verification could bring in a lot of advantages.

We use a metamodel-based automation framework \cite{metamodel} to prove the functional correctness of the design at the IP level, use MSI models to prove the correctness under metastability influence, generate CDC coverage model and later on, use the same MSI and properties setup in simulation-based verification to prove the testcases under metastability influence at SoC level.

\subsection{Sign-off flow for CDC Design}
The CDC design process begins with the implementation of the design unit in RTL based on the architecture specification, using clock-oriented partitioning to mitigate CDC problems. Next, a structural CDC analysis is performed on the RTL code, where basic CDC issues such as missing synchronizers, incorrectly placed synchronizers, glitches induced by combinatorial logic in the synchronizer paths and convergence or divergence errors are identified. It is also important for the designer to provide the CDC constraints that define legal/illegal behaviour premises for structural CDC analysis.  Afterwards, a functional analysis is performed to ensure proper synchronization protocols and prevent CDC issues such as data loss. Often designers may black-box parts of the DUV to simplify the proof process, however, an incorrect black-box may hide crucial CDC related bugs. SystemVerilog assertions generated by the CDC code generator tool are used to verify the CDC constraints, ensuring no incorrect black-boxing of the design. Finally, a formal functional analysis with metastability injection is performed using a comprehensive MSI model to verify the synchronization scheme without metastability-induced errors. This sign-off flow for the CDC design should be followed to produce the CDC-compliant RTL code, and an example of a CDC constraint check (SystemVerilog assertion) generated by the CDC code generator tool is provided in Listing \ref{constraint_check}.
\vspace{0.25cm}
\begin{lstlisting}[language=Verilog, float=h!, caption=Example code for CDC constraint check, basicstyle=\ttfamily, label={constraint_check}]
// static signal check with condition
property static_check_sig;
    (!$stable(sig)
  |->
    (!sig_static_cond));
endproperty
assert property(@(posedge clk) static_check_sig);
\end{lstlisting}

\subsection{Sign-off flow for CDC Verification}
The verification sign-off flow employs a formal verification setup to verify the design at the IP level, followed by simulation-based testing at the SoC level. Fig.~\ref{methodology_flow} depicts the required steps for the CDC verification closure of multi-clock asynchronous designs. The flow includes both formal-based methods to verify the design at the Intellectual Property (IP) level and at the same time, reuse of IP level setup at the SoC level to perform CDC verification with MSI.

\begin{figure}[h!]
\centering
  \includegraphics [width=0.7\textwidth] {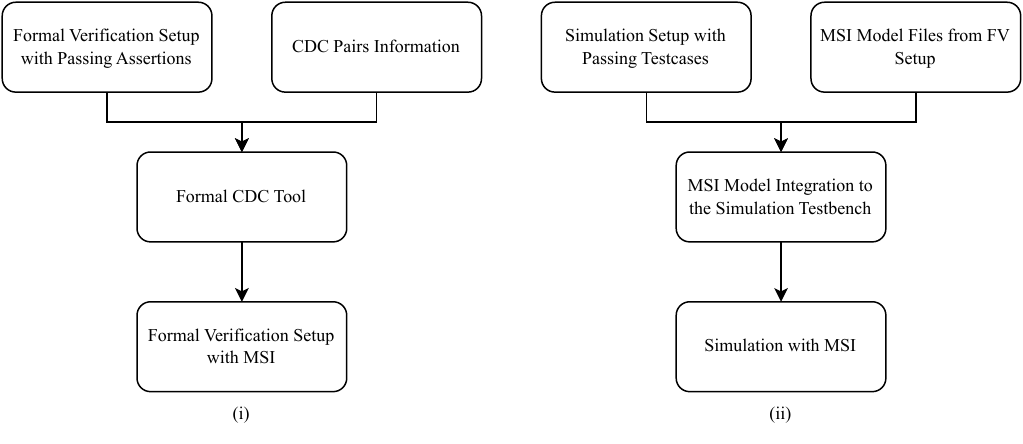}
\caption{Metastability injection flow in (i) formal verification setup and (ii) simulation setup}
\label{formal_sim_flow}
\end{figure}

\subsubsection{Formal Verification Setup}
The CDC verification flow begins with creating a plan based on the design specification, including functional features to address. SystemVerilog assertions are written by the verification engineer to verify these features and are complemented by assertions from the CDC code generator based on synchronizer schemes. The formal verification setup checks these assertions using CDC-compliant RTL code. The next step involves validating the assertions with metastability injection to ensure proper functionality despite metastability effects. Finally, the CDC coverage model generated by the CDC generator tool is analyzed to sign off the verification at the IP level. Fig.~\ref{formal_sim_flow} (i) shows the metastability injection flow in the formal verification setup. To enable MSI support, the formal CDC tool, in our case Cadence Jasper needs two inputs: the formal setup with passing assertions and information about CDC pairs in the design. The tool collects information on CDC pairs during structural analysis and does not need the user to explicitly provide this information. The MSI model generated by the tool automatically handles both the setup and hold timing violations for every CDC pair. Since we start with passing assertions, any assertion failure after running MSI enabled formal proofs clearly indicates metastability effects to be the reason for failure.

\subsubsection{Simulation Setup}
The SoC level design is verified using a simulation-based setup. The process starts with a simulation regression to verify functional features at the chip level. Once all regression test cases pass, a metastability injected regression ensures the design works correctly in presence of metastability effect. Additionally, CDC functional checks (SystemVerilog assertions) from the formal verification setup are exported to the simulation to validate the synchronizer protocols in the presence of metastability effects. The CDC coverage model is then used to assess the completeness of CDC verification at the SoC level. Fig.~\ref{formal_sim_flow} (ii) illustrates the metastability injection flow in the simulation setup. The inputs to this flow are a simulation setup with passing test cases and MSI model files exported from the formal verification setup (Cadence Jasper). Integrating the MSI model involves instantiating it in the testbench top file. This represents the process of metastability injection in simulation, where we used Cadence Xcelium as the tool of choice.

\subsection{CDC Coverage Model}
\begin{figure}[h!]
\centering
  \includegraphics [width=0.35\textwidth] {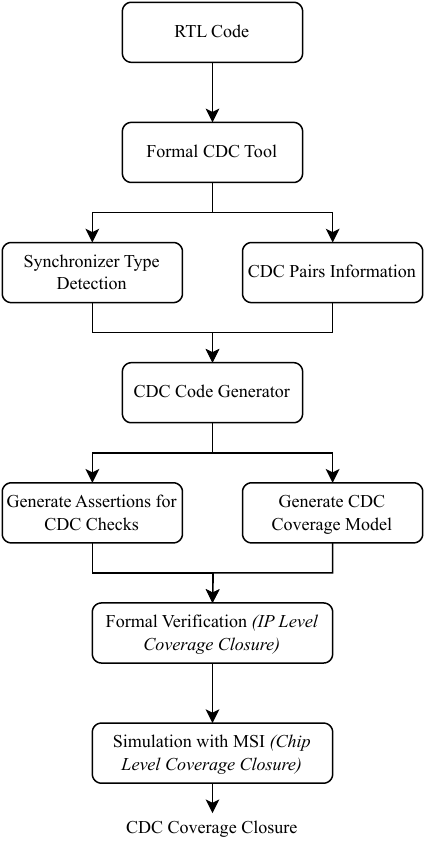}
\caption{CDC coverage model}
\label{cov_model}
\end{figure}

A CDC coverage model is developed to analyze the completeness of CDC verification as illustrated in Fig.~\ref{cov_model}

The coverage model consists of two levels of coverage:

\begin{itemize}
    \item IP level coverage - indicates the CDC pairs coverage at the block level and is evaluated using a formal verification setup.
    \item SoC level coverage - indicates the CDC pairs coverage at the chip level and is evaluated using a simulation-based setup with MSI.
\end{itemize}

The CDC code generator tool requires two inputs: synchronizer types and CDC pairs information from the CDC-compliant RTL code. The tool generates the CDC coverage model (Listing \ref{cov_model_sv}) for formal verification. The developed coverage model analyzes the coverage of the following cases for all CDC pairs:

\begin{itemize}
    \item Setup time violation resulting in a logic 0 at the destination flop
    \item Setup time violation resulting in a logic 1 at the destination flop
    \item Hold time violation resulting in a logic 0 at the destination flop
    \item Hold time violation resulting in a logic 1 at the destination flop
\end{itemize}
\vspace{0.25cm}
\begin{lstlisting}[language=Verilog, float=h!, caption=Generated CDC coverage model, basicstyle=\ttfamily, label={cov_model_sv}]
// CDC coverage model
module cdc_coverage;
  // CDC pair signals extracted from the CDC generator tool
  logic sig_src_1;
  logic sig_dest_1;
  // coverpoints are sampled wrt clock 'clk'
  covergroup cg @(posedge clk);
    // coverage for CDC pair sig_src_1 and sig_dest_1
    cp_src_1      : coverpoint sig_src_1;
    cp_dest_1     : coverpoint sig_dest_1;
    cx_cdc_pair_1 : cross cp_src_1, cp_dest_1;
  endgroup
  // instance for CDC coverage
  cg cdc_cover_inst = new();
endmodule
\end{lstlisting}

\subsection{CDC Code Generator}
\begin{figure}[h!]
\centering
  \includegraphics [width=0.55\textwidth] {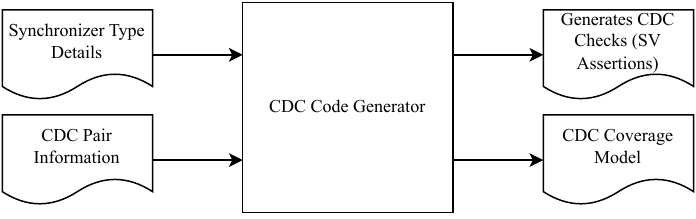}
\caption{CDC code generator}
\label{code_generator}
\end{figure}

The CDC code generator tool is developed to generate both the CDC checks (SystemVerilog assertions) and the CDC coverage model. As shown in Fig.~\ref{code_generator}, the CDC code generator requires two inputs to generate the CDC checks and the coverage model. The first input is the details of the synchronizers present in the design. The Jasper formal CDC app is used to extract this information about synchronizers from the design. The second input is the CDC pair information, which is also extracted similarly from the Jasper CDC app.

The CDC code generator tool produces SystemVerilog assertions for various CDC checks, including functional analysis of synchronizer schemes, clock gating checks, and signal configuration checks. It supports standard synchronizer schemes (2-DFF, data path, pulse type, mux-based, multi-bit) and some custom synchronizers. The tool also generates signal configuration checks to validate CDC constraints set during structural CDC analysis. Additionally, it creates a CDC coverage model to analyze the coverage of CDC pairs in the design. An example of the SystemVerilog assertions generated by the tool is shown in the Listing \ref{cdc_assertion_sv}.
\vspace{0.25cm}
\begin{lstlisting}[language=Verilog, float=h!, caption=Generated assertions by CDC code generator, basicstyle=\ttfamily, label={cdc_assertion_sv}]
// pulse synchronizer checks
// assumptions/constraints
property pulse_input_width_pulse_i;
  @(posedge prim_clk_i)
    disable iff (!prim_reset_n_i)
    $rose(pulse_i)
  |->
    ##1 !(pulse_i)[*2] ##1 (pulse_i) ##1 !(pulse_i) ##1 !(pulse_i);
endproperty
assume property(pulse_input_width_pulse_i);

// 1. output pulse width check
property pulse_output_width_check_pulse_o;
  @(posedge sec_clk_i) disable iff (!sec_reset_n_i)
    pulse_o |=> !pulse_o;
endproperty
assert property(pulse_output_width_check_pulse_o);

// 2. pulse transfer check
property pulse_transfer_check_pulse_i;
  @(posedge prim_clk_i) disable iff (!prim_reset_n_i || !sec_reset_n_i)
    pulse_i |-> @(posedge sec_clk_i) ##[0:3] oulse_o ##1 !pulse_o;
endproperty
assert property(pulse_transfer_check_pulse_i);
\end{lstlisting}

\begin{figure}[h!]
\centering
  \includegraphics [width=0.7\textwidth] {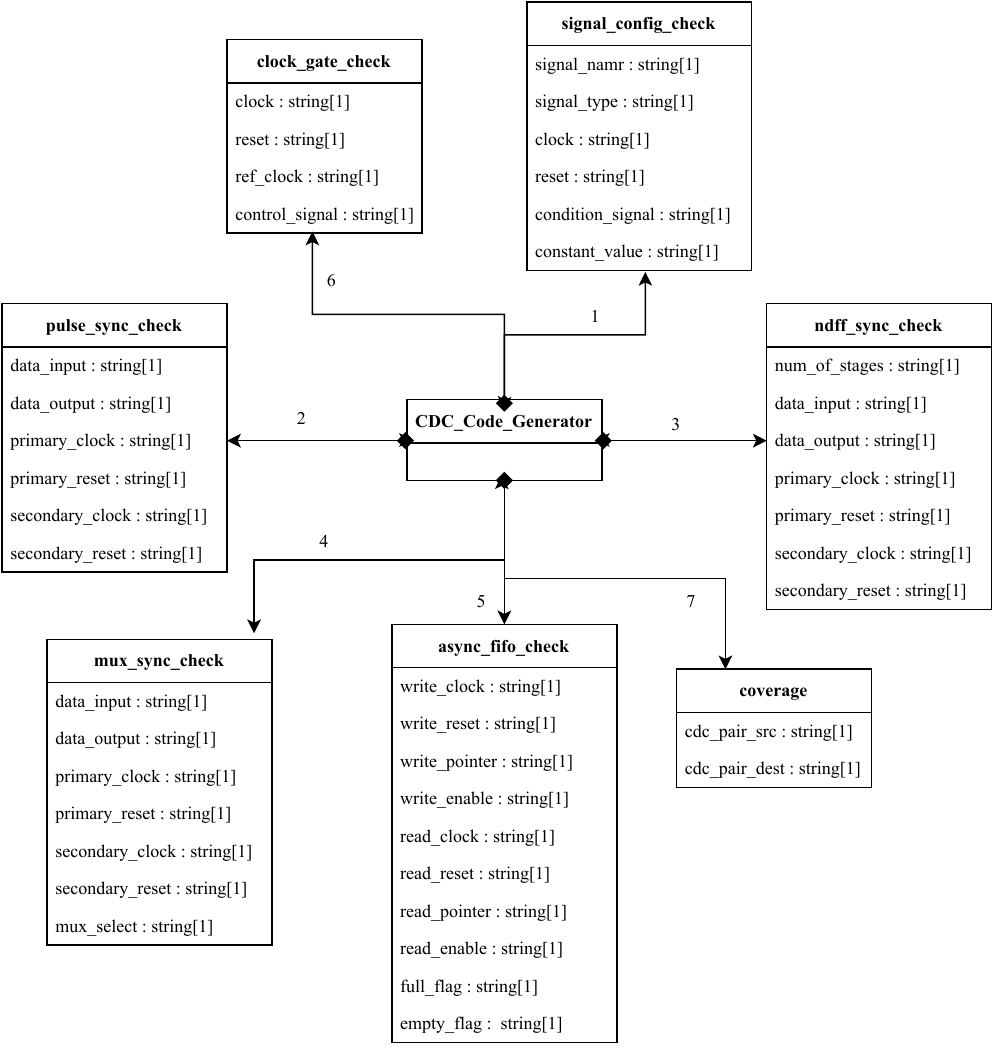}
\caption{UML class diagram of the CDC code generator}
\label{uml}
\end{figure}

The code generator tool is developed using the metamodel-based automation framework \cite{metamodel}. Fig.~\ref{uml} displays the UML representation of the metamodel, illustrating the relationships between its classes. The generator's metamodel includes separate classes for each synchronizer type and a class responsible for generating the coverage model.

The metamodel of the generator consists of the following models with distinct functionalities:

\begin{itemize}
    \item signal\_config\_check - generates checks for signal configuration and CDC constraints specified by the designer.
    \item pulse\_sync\_check - generates checks for pulse synchronizers
    \item ndff\_sync\_check - generates checks for NDFF synchronizers, where ’N’ denotes the number of flip-flops used
    \item mux\_sync\_check - generates checks for multiplexer-based synchronizers
    \item async\_fifo\_check - generates checks for asynchronous FIFO
    \item clock\_gate\_check - generates checks for clock gating feature
    \item coverage - generates CDC coverage model
\end{itemize}

After generating the metamodel, the framework parses it to create Python-based Application Programming Interfaces (APIs) and Graphical User Interface (GUI) files of the tool. The GUI of the 'CDC code generator' is then invoked to set the values of the class attributes and finally, the specification is saved in an XML format. The CDC checks and coverage model are finally generated in two separate files afterwards.

\section{Results}
\label{results}
The methodology successfully verified the design units for CDC issues at the IP and chip levels. Formal CDC verification with metastability injection, along with CDC coverage analysis, was done successfully for the IP level CDC verification. Similarly, simulation with metastability injection and CDC coverage analysis was performed successfully for the chip-level CDC verification.

The application of the verification methodology developed on different design units such as common cells, ethernet controller IP, and an SPI block working in different clock domains unveiled several hard-to-find bugs. It was successful in finding many RTL design bugs and testbench bugs. The types of bugs identified with this verification methodology are shown in Table \ref{bugs}.

\begin{table}[h!]
\centering
\begin{tabular}{|c|c|l|}
\hline
\rowcolor[HTML]{EFEFEF}
\textbf{Bug Type} & \textbf{Analysis Used} & \multicolumn{1}{c|}{\cellcolor[HTML]{EFEFEF}\textbf{Bug Description}}    \\ \hline
RTL bug           & Structural analysis    & Missing synchronizer for CDC signal                                      \\ \hline
RTL bug           & Structural analysis    & Missing synchronizer for RDC signal                                      \\ \hline
RTL bug           & Structural analysis    & Combinational logic on the CDC path                                      \\ \hline
RTL bug           & Structural analysis    & Combinational logic on the RDC path                                      \\ \hline
RTL bug           & Structural analysis    & Reset signal converged before reaching the destination unit              \\ \hline
RTL bug           & Functional analysis    & Wrong signal configuration (signal wasn't static)                        \\ \hline
RTL bug           & Functional analysis    & Signal not stable enough to be captured correctly by destination clock                     \\ \hline
Testbench bug     & Functional analysis    & Input pulse was more than 1 cycle wide (pulse synchronizer)              \\ \hline
Testbench bug     & Functional analysis    & Data loss because of incorrect clock frequencies                         \\ \hline
Testbench bug     & Metastability analysis & Assertion didn’t handle the extra delay due to metastability propagation \\ \hline
Formal tool bug   & MSI model generation   & MSI model was not getting generated for the simulation                   \\ \hline
\end{tabular}
\caption{Types of CDC bugs detected}
\label{bugs}
\end{table}

\begin{figure}[h!]
\centering
  \includegraphics [width=\textwidth] {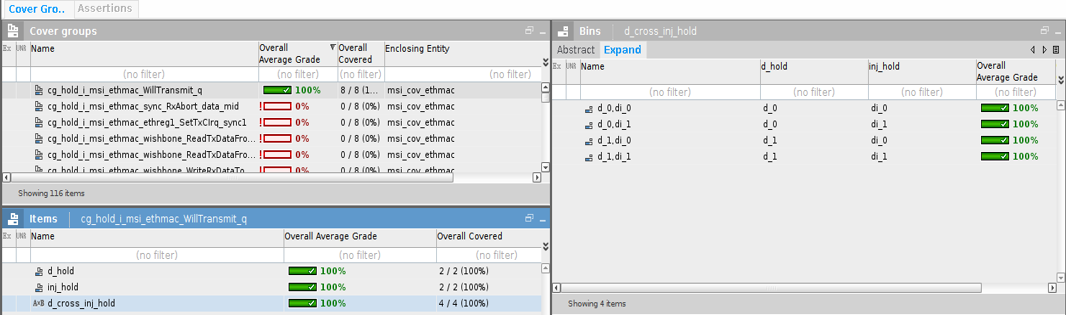}
\caption{CDC pairs coverage}
\label{cdc_cov}
\end{figure}

In addition to finding the above-mentioned bugs, the methodology also analyzed the CDC pairs coverage using the developed CDC coverage model and the one generated by Cadence Jasper \cite{cadence_cdc}. As shown in Fig.~\ref{cdc_cov}, the methodology was successful in finding many CDC pairs with extremely poor coverage. This shows that the developed methodology is efficient in finding potential CDC issues.

\section{Conclusion}
\label{conclusion}
In this paper, we presented a verification methodology for multi-clock SoC architectures to address CDC issues. The methodology bridges the gap between structural and functional verification and employs simulation and formal verification methods, as well as metastability aware verification, to detect all CDC design bugs. The methodology consists of five stages of CDC analysis: structural, functional, metastability, simulation with metastability injection, and CDC coverage analysis. The methodology successfully detected hard-to-find corner case CDC-related issues and reduced the designer's effort in generating the jitter model by offering a comprehensive metastability injection model. In general, the methodology is exhaustive, efficient, measurable, and scalable.

\printbibliography[heading=bibintoc]\label{sec:bibliography}%

\end{document}